\documentstyle[aps,epsf,epsfig,rotate]{revtex}
\newcommand \ee{\end{equation}}
\newcommand \be{\begin{equation}}
\def\bi{\bibitem}
\newcommand \bea {\begin{eqnarray} \nonumber }
\newcommand \eea {\end{eqnarray}}
\def\bphi{{\mbox{$\vec{\phi}$}}}
 \def\(({\left(}
 \def\)){\right)}
\def\[[{\left[}
\def\]]{\right]}
\def\bi{\bibitem}

\def\l{\lambda}

\begin{document}

\title{Aging classification in glassy dynamics}
\author{A. Barrat, R. Burioni and M. M\'ezard}

\address{Laboratoire de Physique Th\'eorique de l'Ecole
Normale Sup\'erieure\footnote{Unit\'e propre du CNRS, associ\'ee \`a
 l'Ecole Normale Sup\'erieure et \`a l'Universit\'e de
Paris Sud}, 24 rue Lhomond, 75231 Paris Cedex 05, France}
\date{\today}
\maketitle
\begin{abstract}
We study the out of equilibrium dynamics of several models
exhibiting aging. We attempt at identifying various types of
aging systems using a phase space point of view:
we introduce a trial classification, based on the overlap
between two replicas of a system,
which evolve together until a certain waiting time, and are
then totally decoupled. We investigate in this way two
types of systems, domain
growth problems and spin glasses, and we show that
they behave differently.
\end{abstract}
\vskip 1cm
\begin{center}
LPTENS preprint 95/44
\end{center}


Submitted to: J. Physics A

\newpage
\section{Introduction}

The dynamics of spin-glasses and other disordered systems exhibits
a very much studied phenomenon known as ``aging'':
the behaviour of the system depends on its history,
and experiments show a typical out of equilibrium
regime on all (accessible) time scales \cite{exp}. In the simplest
case one quenches the system into its low temperature
phase at time $t=0$, and the dynamics of the system
depends on its age, i.e. the time elapsed since the quench.
This type of behavior can be studied for example by looking
at the correlation function of some local observable
$O(t)$, $C(t,t')=<O(t)O(t')>$, or at
the response of such an observable to a change in a conjugated external field
$h(t')$: $r(t,t') = <\frac{\partial O(t)}{\partial h(t')}>$.
While in the usual equilibrium behaviour
these two-times quantities obey time translational invariance (TTI)
($C(t,t')=C(t-t')$, $r(t,t') =r(t-t')$)
and fluctuation-dissipation theorem (FDT) relating
correlation and response, one frequently observes in off equilibrium
dynamics a dependence in $C(t,t') \simeq t^{-\alpha} C(t'/t)$,
which is referred to as an aging behaviour, and violation of FDT.

This kind of aging behaviour is not restricted to spin glasses:
the persistence of out of equilibrium effects
even after very long times has been observed in
many other systems, either experimental systems \cite{exp2},
or in computer simulations \cite{simul}.
In some cases aging could be studied analytically
\cite{CuKu,franzmezard,cukusk,cukupar,Ritort,BM}.

The kind of loose definition of aging that we have used so far seems
to be ubiquitous and to hide a variety of very distinct physical
situations. While the mean field spin glass is known to be a complicated
system with a rough free energy landscape with many metastable
states, aging also occurs in much simpler problems like the
random walk \cite{cukupar}, the coarsening of domain walls in a ferromagnet
quenched below its critical temperature \cite{bray},
or some problems with purely entropic barriers \cite{Ritort,BM},
all problems in which the free energy landscape seems to be very simple.

It is interesting to find a way to
distinguish between these different types of aging, and this paper
wants to make some step towards such a classification.
A first classification of aging has already been proposed in the
litterature \cite{franzmezard}. In the mean field dynamics of
spin glasses, it has been shown that the response function exhibits
an anomaly in the low temperature phase \cite{sompzip,horner}.
While it looks mysterious in the framework of equilibrium dynamics,
this anomaly is well understood if one studies off equilibrium
dynamics \cite{franzmezard}. The anomaly is there defined as
\be
\bar{\chi} = \lim_{t \to \infty} \int_0^t \ dt' \ r(t,t')
- \int_0^\infty \ \lim_{t_w \to \infty} r(t_w + \tau, t_w) d\tau
\ee
It measures the difference between the susceptibility of  the system
at large times and the susceptibility of an hypothetical system which
would be at equilibrium.
A non-zero anomaly shows the existence of a long term memory of the system
to some perturbations occuring at any time. Systems with such an anomaly
certainly exhibit strong aging effects.

In  spite of its nice mathematical structure, the anomaly is in general
not easy to control and compute (analytically or numerically). In this paper we
want to propose another tool for the classification of aging.
We shall use an overlap $Q_{t_w}(t_w+t,t_w+t)$ between two
identical copies of the system, which are constrained to evolve
from the same initial configuration and with the same thermal
noise between the initial quench and a time $t_w$, and
then evolve with different realizations of the thermal noise between $t_w$
and $t_w+t$. This quantity was introduced in \cite{andrea},
and in a study of the
aging dynamics of the spherical spin glass by Cugliandolo and Dean
\cite{cugldean} (slightly different objects involving two copies
of the system evolving with the same noise had
also been studied before \cite{derrida}).
We argue that the asymptotic value of this overlap in the
double limit $lim_{t_w \to \infty} lim_{t \to \infty}Q_{t_w}(t_w
+t,t_w+t)$ distinguishes
between different types of aging. In a first class of systems
the limit is finite (equal to the Edwards Anderson parameter $q_{EA}$
in the cases we have studied so far). This class which we call of type I
includes the models with coarsening of domain walls:
we show it explicitly hereafter
in the case of the $O(n)$ model with $n \to \infty$, and within some
widely used assumptions for the domain growth in the non conserved
scalar order parameter case. The second class, aging systems of type II,
contains the spin glass like problems with complicated free
energy landscapes, and we study explicitely the p spin spherical
models or the zero dimensional version of the manifolds in random potential.
For this class, the limit $lim_{t_w \to \infty} lim_{t \to \infty}$
of the overlap is equal to the minimum possible overlap (i.e.
zero for the p spin spherical model -with $p > 2$- and
$q_0$ for the zero dimensional manifolds).

Besides suggesting a first (rough) classification, this overlap
function may turn to give some intuitive ideas about the
energy landscape in which the system evolves, and its
complexity.
For example, if we think of a system falling down a ``gutter'', it
is clear that it will continuously go away from its position at
$t_{w}$ (the correlation function decreases to zero), but
two copies separated at $t_{w}$ will not be able to separate
indefinitely, and the overlap will have a limit at long
times which can depend on $t_{w}$: as $t_w$ grows, the system
is going closer and closer to equilibrium. Type I systems
seem to have such a behaviour.

On the contrary, a rugged landscape with many bifurcations,
and many different paths,
will allow two copies to really move away from one another,
so the overlap will decay to its minimum possible value, for any
finite $t_{w}$:
the distance between the two replicas becomes the largest possible one.

The classification induced by the asymptotic value of the overlap function
might coincide with the one quoted above, using the anomaly of the response.
Indeed all type II systems that we study are known to possess a non
vanishing anomaly. In type I systems, the anomaly has been computed so
far only in the $O(n)$ model with $n \to \infty$, where it does vanish
(or equivalently for the $p=2$ spherical model\cite{cugldean}).
The existence of a general relation between these two criterions remains
to be studied. At an intuitive level it may look plausible: if one
thinks of a type I system as evolving in a phase space gutter, it should
not have a long term memory of a perturbation. On the other hand,
a type II system evolving in a rugged landscape will be continuously
bifurcating and a change of direction will be remembered at long times.
As we are aware that such intuitive arguments can be very misleading,
we just mention them here as a motivation to further studies of the
response anomaly in various aging systems.

The paper is organized as follows: in section II,
we define the dynamics and various quantities we study, and present the general
features of equilibrium dynamics. Section III is devoted
to the study of various problems of domain-growth, with
analytical and numerical results. Type II systems are studied in
section IV, where we analyse in particular the behaviour
of the zero dimensional version of the random manifold
problem  and of Bouchaud's model of phase space traps \cite{Bouchaud1,BouDean}.
The last section contains our conclusions.

\section{Definitions, equilibrium dynamics}

We consider systems described by a field $\phi(\mbox{\boldmath $x$})$
in a d dimensional space (we shall also consider spin systems, with
obvious generalizations of the definitions).
Given a Hamiltonian $H[\phi]= \int d^d x \cal{H}(\phi
(\mbox{\boldmath $x$}))$, we assume a Langevin dynamics
at temperature $T$:
\be
{\partial \phi(\mbox{\boldmath $x$},t) \over \partial t} =
 - {\partial H \over \partial
\phi(\mbox{\boldmath $x$},t)} + \eta(\mbox{\boldmath $x$},t),
\ee
where $\eta$ is a white noise, with $<\eta(\mbox{\boldmath $x$},t)
\eta(\mbox{\boldmath $x'$},t')>
= 2 T \delta^d(\mbox{\boldmath $x$}- \mbox{\boldmath $x'$})\delta(t-t')$
($< >$ means an average over this thermal noise).

The quantities we are mostly interested in are the following:
\begin{itemize}
\item the two-time autocorrelation function $C(t,t')$: it is the mean overlap
between the configurations of the field at times $t$ and $t'$,
\be\label{cc}
C(t,t')= \frac{1}{V}\int d^d x < \phi(\mbox{\boldmath $x$},t)
 \phi(\mbox{\boldmath $x$},t') >;
\ee
\item the response function
\be\label{rr}
r(t,t')=  \frac{1}{V} \int d^d x < \frac{\delta \phi(x,t)}
{\delta \eta(x,t')} >;
\ee
\item the overlap function $Q_{t_w}(t,t')$: the system evolves
during a certain time $t_w$; at $t_w$ a copy is made, and the
two systems obtained, labeled by $^{(1)}$ and $^{(2)}$, evolve
independently; $Q_{t_w}(t,t')$ is then the overlap between the
configuration of one copy at time $t$ and the other at time
$t'$:
\be\label{qq}
Q_{t_w}(t,t')=\frac{1}{V}  \int d^d x < \phi^{(1)}(\mbox{\boldmath $x$},t)
\phi^{(2)}(\mbox{\boldmath $x$},t') >.
\ee
Of course, for $t \leq t_{w}$ or $t' \leq t_{w}$, $Q_{t_w}(t,t')=C(t,t')$.
\end{itemize}

Before turning to out of equilibrium dynamics, let us first
show that the overlap $Q_{t_w}(t,t')$ is simply related to
the correlation in the case of equilibrium dynamics.

If a system is evolving among a set of states,
according to a master equation,
with transitions rates obeying detailed balance,
i.e.:
\bea
\frac{d}{dt}p_{i}(t)=\sum_{j} \ T_{ij} \ p_{j}(t) \nonumber \\
T_{ij} \ p_{j}^{eq} = T_{ji} \ p_{i}^{eq}
\eea
where $p_i(t)$ is the probability of being in state
$i$ at time $t$, the formal solution is
\be
p_{i}(t)= \sum_{j} < i | e^{{\cal{T}}t} | j >  p_{j}(0),
\ee
where $< i | e^{{\cal{T}} t} | j >$ are the matrix elements of the
evolution operator $e^{{\cal{T}}t}$ ($< i |{\cal{T}} | j >=T_{ij} $).
The detailed balance implies that:
\be
< j | e^{{\cal{T}}t} | i >  p_{i}^{eq}  = < i | e^{{\cal{T}}t} | j >
p_{j}^{eq}
\ee

If we express this property in terms of the overlap between
two replicas evolving in equilibrium dynamics, we obtain (see figure
(1)):
\be
Q_{as}(s,t)=C_{as}(s+t)
\ee
where we have defined $Q_{as}(s,t)= \lim_{t_w \to \infty}
Q_{t_w}(t_w+s,t_w+t)$
and $C_{as}(t) = \lim_{t_w \to \infty} C(t_w+t,t_w)$.
\begin{figure}
\centerline{\hbox{
\epsfig{figure=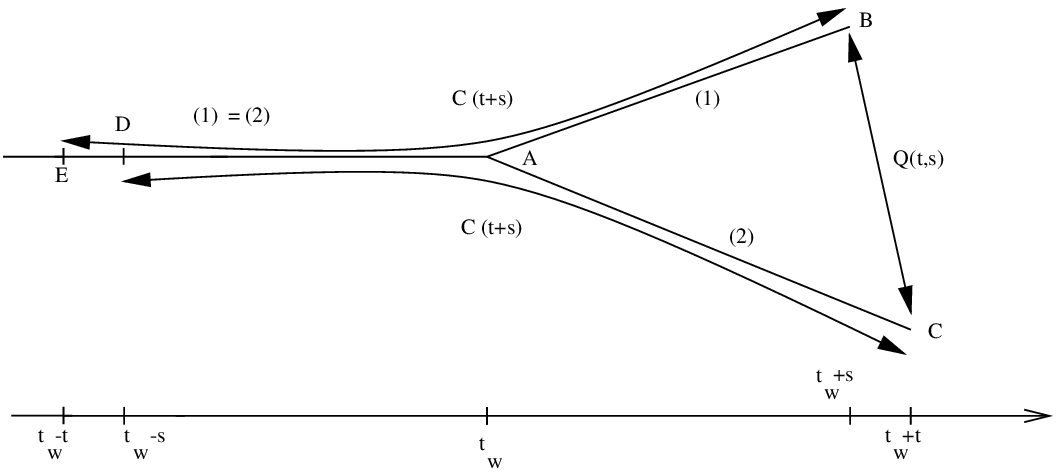}
}}
\caption{
the equilibrium value of the overlap, when $t_w \to \infty$
between the first replica at time
$t_w+s$ (point B) and the second at time $t_w+t$ (point C) is the same
as the overlap between D (time $t_w-s$, before the separation) and
C), or between E (time $t_w-t$) and B
}
\end{figure}

There exist other interesting large time limits in the
problem which exhibit aging. In particular,
the interesting
property of weak-ergodicity breaking \cite{Bouchaud1},
defined by $\lim_{t\to \infty} C(t_w+t,t_w)=0$,
expresses the fact that such system never reach equilibrium. In the
following we will therefore be interested in the
function
\be
S(t_w) \equiv \lim_{t\to \infty} Q_{t_w}(t_w+t,t_w+t)
\ee
 and in particular in its large $t_w$ limit
$S_\infty = \lim_{t_w \to \infty}\lim_{t\to \infty} Q_{t_w}(t_w+t,t_w+t)$.
We shall show that this limit
 depends on the type of system one considers and allows for
a distinction of various aging types.

\section{Domain-growth processes}

A phenomenon which is often considered as
the typical example of an out of equilibrium
dynamical evolution is the phase ordering kinetics \cite{bray,gunton}.
It is the domain growth process for
an infinite system with different low temperature ordered phases,
suddenly quenched  from
a disordered high temperature region into an
unstable state at low temperature. Here we shall keep to the
dynamical evolution of systems with a
non conserved order parameter \cite{bray,gunton}.

The system we study is described by a $n$ component vector field
$\phi(\mbox{\boldmath $x$},t)$ representing the density of
magnetization at the point {\boldmath $x$} of
a $d$ dimensional space, as a function of time.
The system is prepared at high temperature, where
$<\phi>~ = 0$ and then rapidly quenched a $t=0$ in a low temperature
region, where there are more than one energetically favourable states
with $<\phi> \ne 0$. This situation is well described by
a typical coarse grained free energy:
\begin{equation}
F  = \int {d^d\mbox{\boldmath $x$}~[~ {1\over 2}(\bf{\nabla}\phi)^2 +
V(\phi)]},\label{enlib}
\end{equation}
where the first term represents the energy cost of an interface
between two different phases and  $V[\phi]$ is a potential
with minima at different values of $\phi$.
The state with $<\phi>=0$ is unstable at low temperature, so
the system evolves by forming
larger and larger domains of a single phase;
at late stage of growth, the typical pattern of
domains is  self similar and the
characteristic size of a domain is $L(t)$. This evolution
can be studied for instance through a Langevin
dynamics with thermal noise $\eta$:
\be\label{lang}
{ d\phi \over dt} = \bf{\nabla}^2 \phi - V'(\phi) + \eta
\ee
Equilibrium is not
achieved  until $L(t)$ reaches the size of the sample.
In an infinite sample one thus observes an aging behaviour in
the correlation function. Roughly speaking the system
remembers its age $t_w$ through the value of its typical domain size
$L(t)$.

\subsection{The {\boldmath $O(n)$} model}

Interestingly enough, one of the few exactly solved
models of coarsening \cite{bray}, namely the case of the $O(n)$ model with
$n$ large and a constraint $\phi^2=n$,  is also related to a
problem which looks like a spin glass system.
Indeed consider the following spin glass hamiltonian
\be
H=-\sum_{ij} J_{ij} s_i s_j
\ee
where $s_i$ are real spins with a spherical constraint
$\sum_i s_i^2 =n$, and $J_{ij}$ are random couplings.
This model is usually called the ($p=2$)
spherical spin glass \cite{ktj}. Its Langevin dynamics,
\be
{d s_i \over dt }= \sum_j J_{ij} s_j (t) -z(t) s_i(t) + \eta_i(t)
\ee
where $z(t)$ is a Lagrange parameter inforcing the spherical constraint,
can also be written in the basis where the $J_{ij}$ matrix is diagonal:
\be
{d s_\l  \over dt }= (\l - z(t))s_\l (t) + \eta_\l (t)
\ee
Then the dynamical equation reduces to the one of the $O(n)$
model in Fourier space with $\l = -k^2$.
The only important piece of information
on the $J$ matrix is the behaviour of its spectrum near its
largest eigenvalue $\l^*$. The case of a square root singularity
as for instance the Wigner law is equivalent to a $d=3$ coarsening
problem. Clearly  the spherical spin glass does not really have a
spin glass like behaviour (this has been known for long in the statics
\cite{ktj}); however it exhibits an aging dynamics, which is related to
the growth of the correlation length in the O(n) model.
 Recently Cugliandolo and Dean \cite{cugldean}
have performed a detailed computation of the dynamics  of
 this problem, in which they computed all the relevant quantities
 of interest for our
discussion. They found aging in the correlation function,
but the anomaly of the response as defined in (1)
vanishes, which shows that the memory effects are weak.
besides, the overlap limit $S(t_w)=\lim_{t \to \infty} Q(t_w+t,t_w+t)$
is a continuous function of $t_w$, taking values between
$q_{EA}^2$ and $q_{EA}$, with $S_\infty =q_{EA}$, and, of course,
$lim_{t_w \to \infty} Q(t_w+t,t_w+t') =C_{as}(t+t')$.

Physically, the system evolves in time through two processes: a thermal
noise which affects evenly all the component $s_\lambda$, superimposed
onto a deterministic evolution which amounts to reinforcing the
eigenmodes closer to the $\lambda^*$, and
this deterministic part is dominant over the thermal noise.

The fact that this system has the property
$S_\infty  =q_{EA}$
together with the weak ergodicity breaking property
$lim_{t_w \to \infty} lim_{t \to \infty} C(t_w+t,t_w)=0$ suggests the
existence of a kind of gutter in phase space:
two replicas, even decoupled, remember forever that they are
evolving in the same canal.

\subsection{Scalar order parameter: analytic study}

We now turn to the domain growth problem in the case of a scalar
order parameter. This problem cannot be solved exactly but we shall use
a well known approximation \cite{ojketc},
recently developped by Bray and Humayun
\cite{BrayHum}. We refer the reader to Bray's review \cite{bray}
for a detailed
presentation of the method. The idea is to take advantage of the
universality of domain growth in the scaling regime:
After an initial regime of fast growth, the order parameter
saturates at the equilibrium value inside a domain and the only way
for the system to further decrease the free energy is the
reduction of the surface
of walls between different domains. Therefore, the dynamical properties
of the system at late stage of growth are given by the motions of the
walls and  in particular by their curvature;
the particular shape of the potential $V[\phi]$, provided
it has well separated minima, is not crucial. If the growth is
influenced by an external field, the difference between the
minima introduced by the field will be the relevant variable.
The universality gives the freedom to choose an appropriate
form for the potential in the free energy, and
also a special form for the thermal noise, which make the analysis
more tractable. Specifically the Langevin equation is replaced
by:
\begin{equation}
{\partial\phi(\mbox{\boldmath $x$},t) \over{\partial t}}
        = \nabla ^2 \phi - V_0' [\phi] + \eta( \mbox{\boldmath $x$},t)
V_1'[\phi]
\label{lang2}
\end{equation}
where $\eta(\mbox{\boldmath $x$},t)$ is the gaussian
 white noise with zero mean and correlator:
\begin{equation}
< \eta(\mbox{\boldmath $x$},t) \eta(\mbox{\boldmath $x$}', t')>= 2T \delta
(\mbox{\boldmath $x$}-\mbox{\boldmath $x$}')\delta( t-t')~.
\end{equation}

The field $\phi(\mbox{\boldmath $x$},t)$ is parametrized by
an auxiliary field $m(\mbox{\boldmath $x$},t)$, through: \begin{equation}
\phi[m] = \phi_0 \left ( {2\over \pi} \right ) ^{1/2}
\int_0^m dx  \exp -(x^2/2) =
\phi_0~~ {\mbox erf} [m/\sqrt{2}]~.
\label{erf}
\end{equation}

With the following choice of the two potentials:
\begin{equation}
V_0[\phi] = {\phi_0^2\over \pi} \exp -2
{}~\left[{\mbox erf}^{-1}\left [{\phi\over \phi_0}\right ]\right]^2,
{}~~~V_1[\phi] = {\phi_0^2\over \sqrt{\pi}}
\exp \sqrt{2} {\mbox erf}^{-1}\left[{\phi\over \phi_0}\right],
\label{vuno}
\end{equation}
the field $m$ satisfies a very simple equation:
\begin{equation}
{\partial m \over{\partial t}} =
         \nabla ^2 m + ( 1 - (\nabla m)^2)m + \eta.
\label{langmfin}
\end{equation}

With the wall profile function (\ref{erf}), the field $m(\mbox{\boldmath
$x$},t)$ measures
the distance of the point {\boldmath $x$} from the interface: at infinite
distance
from the wall, the field $\phi$ saturates to its equilibrium value.
Moreover, the potential $V_0$ has the required two wells shape at the two
equilibrium values. The choice of the potential $V_1$ does not
alter this shape and, as can be seen from (\ref{lang2}) and from the fact
that $V_1'[\phi] = \phi'$, it corresponds to
a thermal noise acting only on the interface. This is an
approximation which is not able, for instance, to reproduce
 a process of nucleation of a bubble. In other words, the
value of $q_{EA}$ in this case remains fixed
to the $T=0$ value, $q_{EA}=\phi_0^2$.
However we expect that this approximation will not affect our main
conclusions concerning the various large time limits of the
overlap.

The physical situation of a rapid quench will be represented
by taking the boundary condition for
$m(\mbox{\boldmath $x$},t)$ to be gaussian with zero mean  and correlator:
\begin{equation}
< m(\mbox{\boldmath $x$},0) m(\mbox{\boldmath $x$}',0 )> = \delta
(\mbox{\boldmath $x$}-\mbox{\boldmath $x$}').
\end{equation}

Equation (\ref{langmfin}) can be solved by neglecting the non linear term
or, more correctly, by taking into account
its mean value.
Let us neglect it in a first approach.
Equation (\ref{langmfin}) can then be solved, giving:
\begin{equation}
m(\mbox{\boldmath $x$},t) = \int_{\vert \mbox{\boldmath $k$}\vert < e^{-1}}
{d^d \mbox{\boldmath $k$}\over 2\pi^d}
 e^{i\mbox{\boldmath $kx$}} \left[
e^{(1-k^2)t} m(\mbox{\boldmath $k$},0) +
\int_0^t dt'  e^{(1-k^2)(t-t')}
 \eta(\mbox{\boldmath $k$},t')
 \right],
\label{solm}
\end{equation}
where $e$ is a cut off given by the width of the interface.
The linearity of the equation and the independence
of the boundary condition and the noise
preserve the gaussian character of the probability distribution for
the field $m$. Mean values of functions of the field $\phi$
can be computed
in term of the evolution of the first and second moment for the
gaussian distribution of $m$.
To compute the correlation function $C(\tau +t_w ,t_w)$
we introduce two fields
$m_1=m(\mbox{\boldmath $x$},\tau + t_w)$ and $m_2=m(\mbox{\boldmath $x$},t_w)$.
When computing    the
overlap $Q(\tau + t_w,\tau + t_w)$ the fields $m_1$ and $m_2$ denote
respectively
$m^{(1)}(\mbox{\boldmath $x$},\tau + t_w)$ and $m^{(2)}(\mbox{\boldmath
$x$},\tau + t_w)$. In both cases the joint distribution of $m_1$
and $m_2$ is a gaussian $P(x_1,x_2)$, which we parametrize as:

\begin{equation}
P(x_1,x_2) = {\gamma \over 2\pi \sqrt{\sigma_1 \sigma_2}}
       exp \left[ -{\gamma^2 \over 2}
\left( {x_1^2 \over\sigma_1} + {x_2^2 \over\sigma_2} -
{2fx_1 x_2 \over \sqrt{\sigma_1 \sigma_2}}   \right) \right]
\label{protot}
\end{equation}
with $\sigma_i=<m_i^2>$, $c_{12}=<m_1 m_2>$ and
$f={c_{12} \over \sqrt{\sigma_1 \sigma_2}}$, $\gamma={1 \over{\sqrt{1-f^2}}}$.
In terms of this distribution the correlation (or overlap)
is given by:

\begin{equation}
<\phi[m_1] \phi[m_2]> = \int_{-\infty}^{+\infty} dx_1 dx_2 \phi(x_1)
\phi(x_2) P(x_1,x_2)= \phi_0^2~~ {2 \over \pi} \arcsin f.
\label{prob}
\end{equation}
where the function (\ref{erf}) has been replaced by
$\phi[m]=  \phi_0^2~~sign[m]$, a good
approximation in the large time regime.
The calculation therefore reduces to the computation of the parameter $f$ in
the covariance matrix of the probability
distribution $P(x_1,x_2)$, which is easily obtained from (\ref{solm}).
Defining :
\begin{equation}
{\cal F} (a,b) = \int_0^a  d\sigma \ e^{-\sigma}
 \left(1-{\sigma \over b}\right)^{-{d\over2}}
\left({\mbox erf}\left({\sqrt{b-\sigma}\over e}\right)\right)^d
\label{inte}
\end{equation}
we obtain, for $t_w \gg 1$:
$$
C(\tau + t_w,t_w) =  \phi_0^2~~ {2\over\pi}~~ \arcsin \left\{
\left( {4(\tau + t_w)t_w \over{(\tau + 2t_w)^2}} \right)^{d/4} \right.
$$
\begin{equation}
\left .{ 1 + T {\cal F}(2t_w, \tau + 2t_w) \over
{[ 1 + T {\cal F}(2(\tau + t_w),2(\tau + t_w)) ]^{1/2}
 [ 1 + T {\cal F}(2t_w, 2t_w)]^{1/2}}} \right\}
\label{corras}
\end{equation}
and
\begin{equation}
Q(\tau + t_w,\tau + t_w) =  \phi_0^2~~ {2\over\pi}~~ \arcsin \left\{
{ 1 + T {\cal F}(2t_w, 2(\tau + t_w)) \over
{ 1 + T{\cal F}(2(\tau + t_w), 2(\tau + t_w))}}\right\}.
\label{overas}
\end{equation}

The asymptotic behaviors of $Q$ and $C$ are very similar to those
studied above in the domain growth of a $n \rightarrow \infty$ component
order parameter (or $p=2$ spherical model): the asympotic
relation between $Q$ and $C$ at  $t_w \rightarrow
\infty$ with $\tau$ finite is satisfied and,
for fixed $t_w$ and $\tau$ large, $C$ has the limiting behavior
\begin{equation}
C(\tau + t_w,t_w) \sim \left( { t_w \over\tau } \right)^{d/4}
\end{equation}
while $Q$ does not go to zero and the limiting value
$S(t_w)$ is a continuous function of $t_w$, approaching the
equilibrium value $q_{EA}$ as $t_w$ grows: $S_\infty =q_{EA}$.
So within this approximation this coarsening problem falls
into the type I classification.

{\it Note}: when one includes the effect of the
gradient squared term in (\ref{solm}),
treated as an average term (as in \cite{bray}), the result
is similar except for a
change in  the numerical value of the function ${\cal F}$ in (\ref{inte}),
where a term $c \sigma^{-2}$ is present, instead of $\exp -\sigma$.

In order to check this approximate analytic treatment, we have performed
numerical simulations of domain growth in two dimensions, for
 a scalar field evolving with a Langevin equation, and also
for Ising spins on a regular two-dimensional lattice, with Glauber
dynamics \cite{glauber}.

\subsection{Scalar order parameter: numerical studies}

We have simulated the evolution of a scalar field $\phi$
on a two-dimensional square lattice, according to the Langevin
equation (\ref{lang}), with a quartic $V_0$ and
a bold discretization scheme:
\begin{eqnarray}\label{lang_discr}
\nonumber
\phi(i,j,t+1)&=&\phi(i,j,t)
+ ( \phi(i+1,j,t) + \phi(i-1,j,t) + \phi(i,j+1,t) + \phi(i,j-1,t)\\
&  & - 4*\phi(i,j,t) + \phi(i,j,t) - \phi(i,j,t) ^{3})*h + \eta(i,j,t)
\end{eqnarray}
where $\eta$ is a gaussian noise with zero mean and variance
$2Th$, $h$ being the used time-step. We proceed by parallel
updating of the field, and vary the time-step $h$.
At $t=0$, $\phi(i,j)$ are taken
as independent random variables uniformly distributed between $-1$ and $1$.
We let the system evolve during $t_{w}$ according to (\ref{lang_discr}),
make a copy of it, and let the two copies evolve independently, i.e.
with independent thermal noises. We record the correlation of each
of the copies with the system at time $t_{w}$ and the overlap
between the replicas.

We present simulations at fixed temperature:
we record the overlap and the correlation
function for different values of  $t_{w}$,
The linear size of the system was of $200$ sites, and one run
was made with a $400*400$ lattice. Each simulation was made
with three different values of $h$ ($h=.02$,
$.04$ and $.08$), to check that the results did not depend
on the used time-step. We checked besides that
the $\frac{t}{t_{w}}$ scaling is well obeyed for the
correlation function for large enough $t_{w}$; for the
overlap, no such scaling is found.
We plot the overlap at time $t_{w}+t$ versus the
correlation between times $t_{w}$ and $t_{w}+t$.

The $Q$ versus $C$ curves show quite clearly that the overlap, after
a transient regime where it decays faster than the correlation,
has a finite limit as $C$ goes to zero. This limit grows
with $\beta$ and with $t_{w}$.

\begin{figure}
\centerline{\hbox{
\epsfig{figure=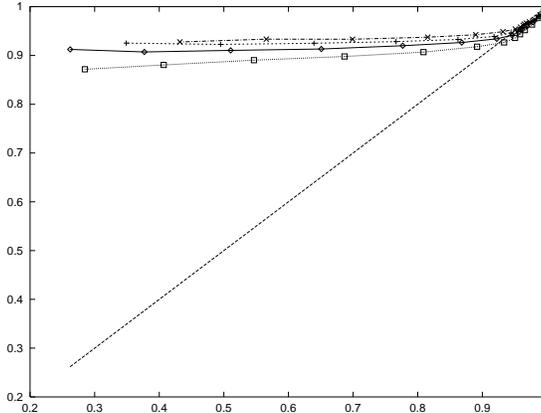,width=5.5cm,angle=-90}
}}
\caption{\small{Overlap $Q(t_{w}+t,t_{w}+t)$ versus
Correlation $C(t_{w}+t,t_{w})$ for the scalar field in two dimensions,
for $\beta=6$ and different waiting times (from bottom to top,
$10$, $20$, $50$ and $180$ MC steps).}}
\end{figure}

This result agrees with the previous analytic study, as far as
the asymptotic behaviour of the overlap and correlation are
concerned. We have also performed simulations of a two-dimensionnal
Ising spin system (with nearest-neighbour ferromagnetic
interactions), with Metropolis dynamics with random updating
\footnote{Notice that the choice of dynamics is important:
as soon as the chosen algorithm is not deterministic at zero
temperature (it is the case for example if we take
Glauber dynamics), the overlap will decrease
to zero even at $T=0$.}: at each sweep through the lattice,
spins are updated in random order, but this
order is the same for both replicas.
The results (see figure 3) agree with those obtained by Langevin
dynamics.

\begin{figure}
\centerline{\hbox{
\epsfig{figure=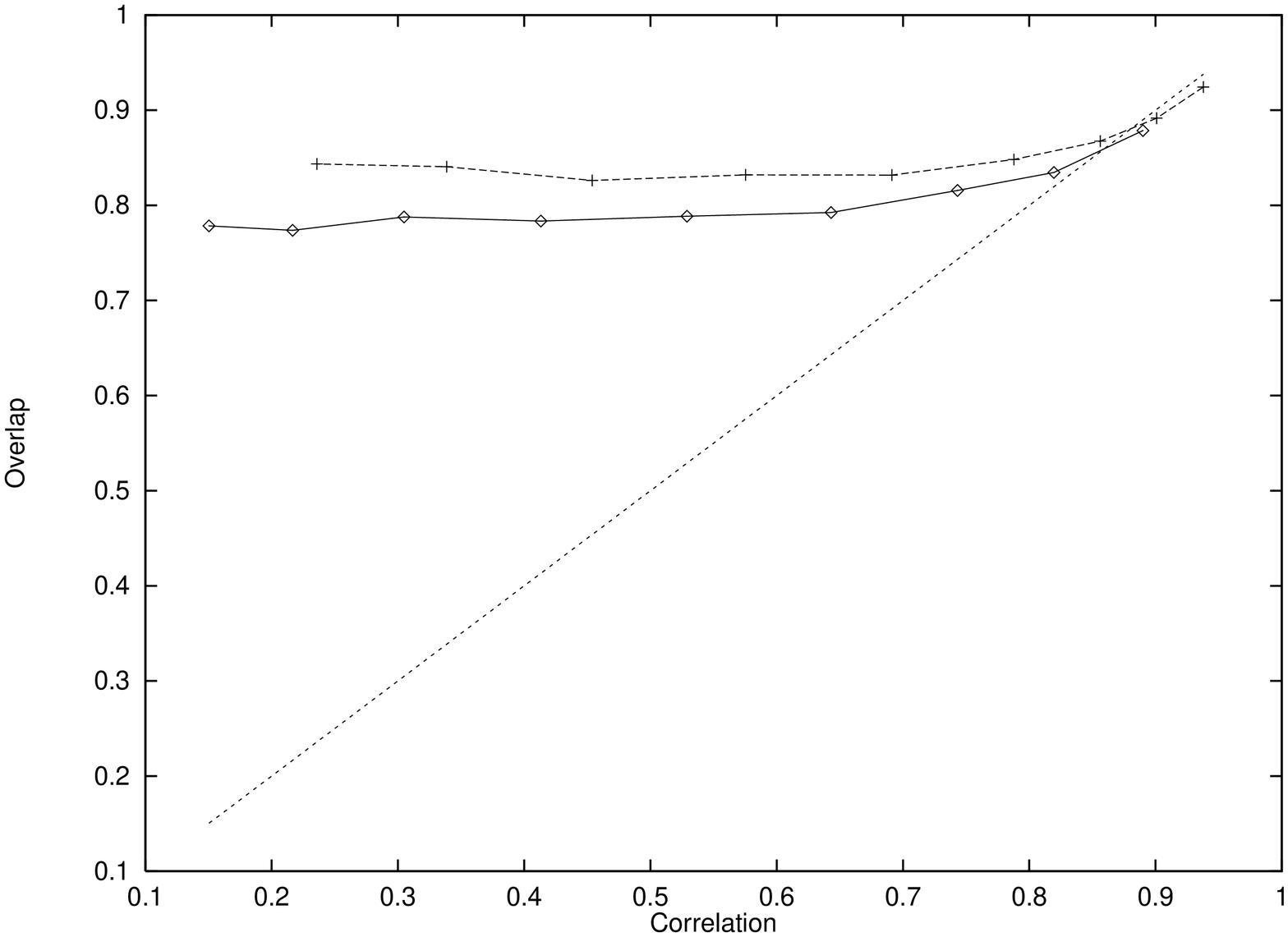,width=5.5cm,angle=-90}
\epsfig{figure=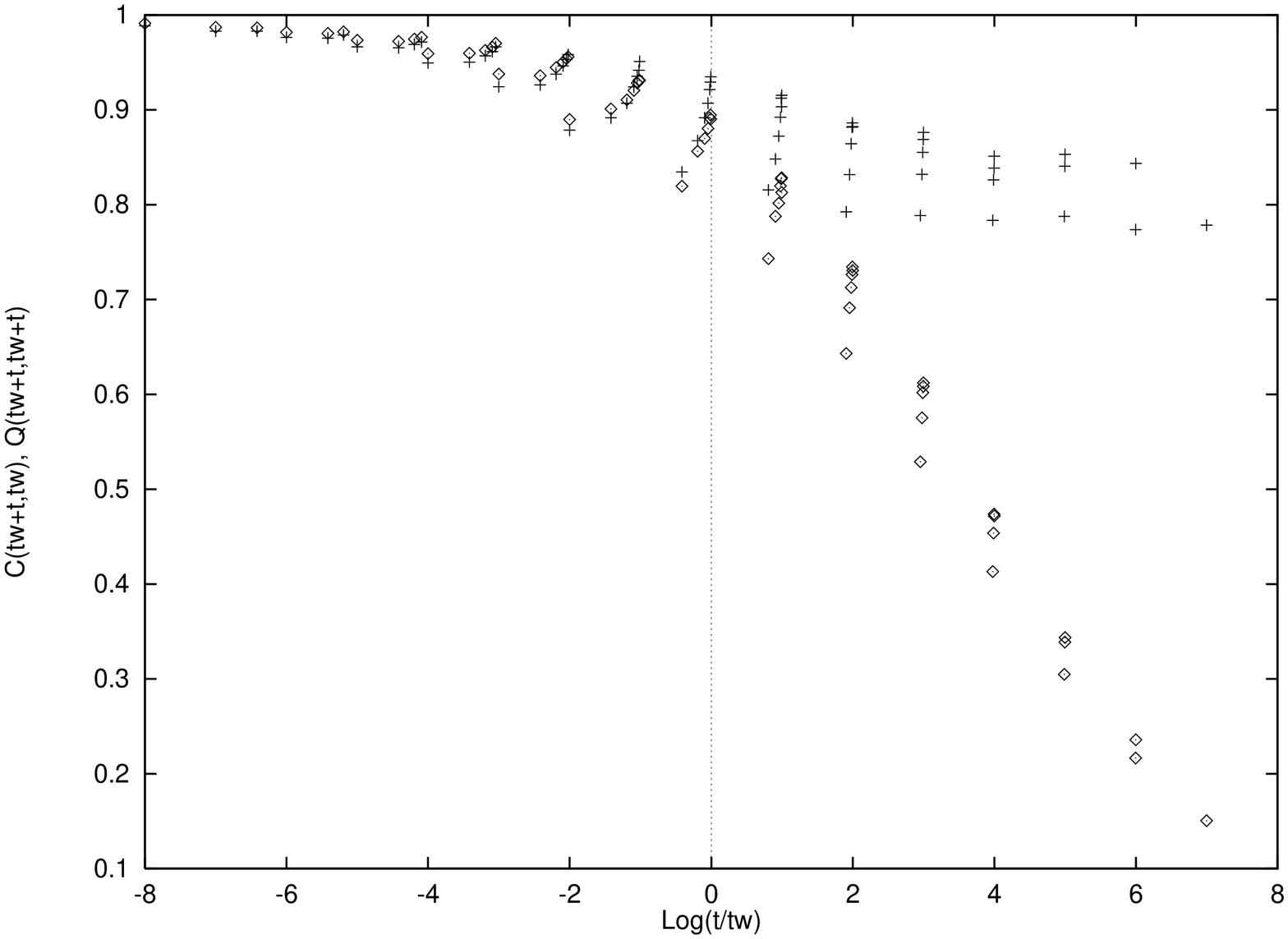,width=5.5cm,angle=-90}
}}
\caption{\small{
Left:Overlap $Q(t_{w}+t,t_{w}+t)$ versus Correlation
$C(t_{w}+t,t_{w})$ for the Ising model in two dimensions,
for $\beta=2$, $t_w=2^3$ (bottom) and $t_w=2^4$ (top); Right:
Correlation and overlap for the Ising model in two dimensions,
for different values of $t_w$ and $t$: $t_w=2^3, 2^4,..., 2^9$, and
$t=2,..., 2^{10}$}}
\end{figure}

{\it Note:} We also made computer simulations for a model introduced
in \cite{sethna}, consisting of an Ising ferromagnet on a cubic
lattice, with weak next-nearest-neighbour antiferromagnetic
couplings; in this model, the growth is slowed from a power-law
to a logarithmic behaviour; nevertheless, we find
for the correlation and overlap functions a similar behaviour
as for the simple ferromagnet.

\subsection{The XY-model in one dimension}

A simple and soluble model where domain growth can be studied
without approximations on the potential
is the $XY$ model in $d=1$ \cite{newman}. Namely, the system has
no phase transition for $T>0$, but at very low temperature
the correlation length
$L_{eq}$ is very large. Then at time scales
where the size of the domains is small
compared to the correlation length, the system presents the
typical non equilibrium features of a multiple phase system.

In this simple model, the order parameter
is a two dimensional vector field $\vec{\phi}(x,t)$ of
fixed length $\vec{\phi}^2 = 1$ and the coarse grained free
energy is:
\begin{equation}
F  = \int dx~ {1\over 2}~
\left ({\partial \vec{\phi} \over \partial x}\right)^2.
\label{enlibxy}
\end{equation}

Using the non linear mapping $ \vec{\phi}(x,t)=
(\cos \theta (x,t), \sin \theta (x,t))$, the Langevin evolution
equation for the field $\theta(x,t)$ can be easily written and explicitly
solved without approximation.
The physical situation of a rapid quench from
a disordered phase
to a very low temperature can be included in the formalism
by taking the boundary condition
$\theta(x,0)$ to be gaussian with zero mean
and correlated at distance $\xi$.
As for the scalar order parameter model,
the linearity of the Langevin equation
preserves the gaussian character of the probability distribution for
the field $\theta$ and the problem can be solved
by computing the time evolution
of the moments.
Let us now consider a quench to a very low temperature,
i.e. a situation where the equilibrium correlation length
$L_{eq}$ is very large, and the time needed for the domains
to reach this size, $t_{eq}$ is also very large: in fact,
$L_{eq} \simeq T^{- \frac{1}{2}}$, and $t_{eq} \simeq T^{-2}$.
In a time regime where  $\tau+ t_w$, $t_w \gg 1$ but $L_{eq}$ is
very large compared
to the size of the domains we have a very simple expression
for the correlation function $C(\tau + t_w,t_w)$
and for the
overlap $Q(\tau + t_w, \tau + t_w)$ of two replicas separated at time $t_w$:
\bea
C(\tau +t_w ,t_w) &=& \exp - {1\over \sqrt{\pi} \xi}
\left[ 2(\tau +2t_w)^{1/2} - [2(\tau+t_w)]^{1/2} - (2t_w)^{1/2} \right.
\nonumber \\
&-& \left. {T\xi\over 2} \{ 2[(\tau + 2t_w)^{1/2}-(\tau)^{1/2}] -
(2(\tau+t_w))^{1/2} - (2t_w)^{1/2} \} \right]
\label{corrlxy}
\eea
and
\begin{equation}
Q(\tau+t_w,\tau+t_w) = \exp -  {1\over \sqrt{\pi} }
T (2\tau)^{1/2}.
\label{qxy}
\end{equation}

Since the size of the domain evolves as
$L(t) \simeq t^{\frac{1}{4}}$, it is clear that there exists
at very low temperatures
a regime with $1 \ll t_w \ll \tau \ll t_{eq}$, where the correlation
has already decayed to zero while the overlap has still a finite
value. Indeed, $C$ decays to zero with a term
in the exponential that does not depend on the temperature,
but the argument of the exponential for $Q(\tau+t_w,\tau+t_w)$
is $L(\tau)^2 / L_{eq}^2$.

\subsection{Conclusion}

The previous study shows that the domain growth processes considered
here are essentially deterministic in nature, and that their phase space
is very simple: we indeed exhibit a time regime
$1 \ll t_w \ll t \ll t_{eq}$ ($t_{eq}$ being the equilibration
time, which is infinite in the true aging problems, but remains
finite in the one-dimensional XY model)
where the system at time $t_w+t$ has
already drifted away from its position in phase space at $t_w$
($C(t_w+t,t_w)$ is very small), while two copies separated
at $t_w$ are still evolving together ($Q(t_w+t,t_w+t)$
is finite). These type I systems are characterized by the
existence of a finite limit $S_\infty=\lim_{t_w \to \infty}
\lim_{t \to \infty} Q(t_w+t,t_w+t)$ (with $S(t_w)=\lim_{t \to \infty}
Q(t_w+t,t_w+t)$ growing continuously towards $q_{EA}$ as
$t_w$ grows).
The system can therefore be thought of as moving
along a gutter in phase space. It is
reasonable to expect that, in these systems, the
influence of the thermal noise will be limited in time,
and there will be no anomaly in the response function
(this has been shown so far only for the O(n) with
$n \to \infty$ model \cite{cugldean}).

\section{Type II models}
\subsection{A particle in a random potential}

We now turn to the study of a well known disordered
mean-field model, where we expect a different kind of
behaviour for the overlap: the toy model described by the
hamiltonian \cite{mmpar,franzmezard}:
\be
H = \frac{1}{2} \mu\sum_\alpha \phi_\alpha^2 + V(\phi_1,...,\phi_N),
\ee
where  $V$ is a gaussian random potential with correlations:
\be
\overline{V(\bphi) V(\bphi')}= -N f\left( {(\bphi-\bphi')^2
\over N} \right) \,
\ee
with:
\be
f(b)={(\theta +b)^{1-\gamma} \over 2(1-\gamma)}.
\ee
This model describes a particle in a random
potential, in $N$ dimensions, but it can also be interpreted
as a spin glass model: the
$\phi_\alpha$ are then soft spins, in a quadratic well
$\frac{1}{2} \mu\sum_\alpha \phi_\alpha^2 $, and they
interact via $V$; the statics has a low temperature
spin glass phase, with continuous replica symmetry breaking
for $\gamma < 1$ (long-range correlations of the disorder) or one-step
replica symmetry breaking for $\gamma > 1$ (short-range correlations).
Slightly
different forms for $f$ allow also to deal with the dynamical
equations of the spherical p-spin\cite{crihornsom,CuKu,cuglledou}
model too. This system is described
by the Hamiltonian
\cite{crisom,CuKu}
\be
\sum_{i_1 < ... < i_p}^{N} J_{i_1 ... i_p} s_{i_1}...s_{i_p}
\ee
with the constraint $\sum_{i=1}^{N} s_i^2 =1 $, and
gaussian distributed random p-spin interactions. It
 can be described by a toy model, with
\be
f(b)= - \frac{1}{2} \left( 1 - \frac{b}{2} \right)^p
\ee
and a small modification of the
 dynamical equations (\ref{em}) and (\ref{ove}),
which amounts to implementing  the spherical constraint by
a time dependent  Lagrange multiplier $\mu(t)$.

To compute the overlap function,
we introduce two replicas $\bphi^{(1)}$ and $\bphi^{(2)}$,
with a Langevin dynamics:
\be
{\partial \phi_\alpha^{(i)}(t) \over \partial t} = - {\partial H \over \partial
\phi_\alpha^{(i)}(t)} + \eta_{\alpha}^{(i)}(t),
\label{langevin}
\ee
where $\eta^{(1)}$ and $\eta^{(2)}$ are two white noises
with $<\eta_\alpha^{(i)} (t) \eta_{\alpha'}^{(i)} (t')>
= 2 T\delta_{\alpha \alpha'}
\delta(t-t')$, and $\eta_\alpha^{(1)}(t) = \eta_\alpha^{(2)}(t)$
if $t \leq t_{w}$. For $t > t_{w}$, $\eta^{(1)}$ and $\eta^{(2)}$ are
uncorrelated.

Using standard field theoretical techniques
\cite{sompzip,kinhorner}, it is now possible to derive the evolution
equations for the correlation and response functions of each replica,
$C^{(1)}(t,t') = C^{(2)}(t,t')=  C(t,t')$, and
$r^{(1)}(t,t') = r^{(2)}(t,t') = r(t,t')$,
and for the overlap $Q_{t_w}(t,t')$, in the large $N$ limit.
These quantities are defined by (\ref{cc}), (\ref{rr}) and (\ref{qq}) and
the corresponding equations are written in appendix A.

The large time limiting values of the correlation
define $\tilde{q}$, $q_0$, and $q_1$ (see \cite{cuglledou}):
\bea
&\lim_{t \to \infty}& C(t,t) = \tilde{q} \\
&\lim_{t \to \infty}& C(t,t') = q_0 \\ \nonumber
\lim_{\tau \to \infty} &\lim_{t \to \infty}& C(t+\tau,t) = q_1
\eea
We will now study the behaviour of the overlap function in different
time regimes.

We first study the regime of asymptotic dynamics
which corresponds to taking
the limit  $t,t'\to\infty$, with $\tau=t-t'$ finite.
We thus obtain the functions
\be
r_{as}(\tau)=\lim_{t'\to\infty}r(t'+\tau,t'), \  \ \
C_{as}(\tau)=\lim_{t'\to\infty}C(t'+\tau,t'),
\ee
\be
Q_{as}(\tau,\tau')=
\lim_{t_w \rightarrow \infty} Q_{t_w}(t_w + \tau,t_w + \tau').
\ee
In this regime, time-translational invariance and fluctuation dissipation
theorem (FDT) are obeyed:
$T r_{as}(\tau)=-{\partial\over \partial \tau}C_{as}(\tau)$. It is well
known that this asymptotic regime is identical to equilibrium dynamics
for systems with long range correlations of the disorder
\cite{kinhorner,franzmezard,cukusk}, but it is different for short
range correlations \cite{crihornsom,kinhorner,CuKu}.
In both cases we have found, as expected from the general
discussion of Sect.II, that
$Q_{as}(\tau,\tau')=C_{as}(\tau+\tau')$.
In particular,
$\lim_{\tau \rightarrow \infty} Q_{as}(\tau,\tau')=
\lim_{\tau' \rightarrow \infty} Q_{as}(\tau,\tau')= q_{1}$.

Let us now consider the aging regime. This regime corresponds to having
time differences, like
$t-t_w$, diverge when $t_w \to \infty$. Here we shall consider the
overlap function $Q_{t_w}(t,t')$ in the ``double aging'' regime where
 $t'-t_w$ also diverges.

There is no full solution of the aging regime in spin glass systems.
What has been proposed so far, in all cases, is an Ansatz about the
behaviour of the correlation or response. The first such proposal,
by Cugliandolo and Kurchan \cite{CuKu}, concerns the case of
the p-spin model. They showed that the dynamical equations can be solved
in the long time regime (where one can neglect the time derivatives
in (\ref{em}) ) by the Ansatz: $C(t,t')={\cal{C}}(t'/t), \
r(t,t')= (x/T) {\cal{C'}}(t'/t)$, or actually by any solution obtained from
this one through a reparametrization of time $t \to h(t)$, with h
an arbitrary increasing function. This solution was subsequently
extended to more complicated problems in which the static solution involves
a full rsb, like the toy model with long range  correlations of the
noise \cite{franzmezard} and the SK model \cite{cukusk}. The
case of the toy model with short range correlations of the noise has
also been studied recently \cite{cuglledou}.  The formalisms
developped in \cite{franzmezard} (non overlapping time domains) and
in \cite{cukusk} (triangular relations) represent the same Ansatz but
look rather different. Here we shall
present the Ansatz  using mainly the former approach,
together with the necessary ingredients for understanding the
correspondence between the two formalisms.

Considering first two time quantities like the correlation or response,
the aging regime corresponds to sending $t$ and $t'$ both to infinity,
the difference $t-t'$ being itself divergent in this limit. The  dynamical
equations can be solved in this limit (up to a time
reparametrization), neglecting the time derivatives.
We consider non overlapping domains in the $(t,t')$ plane:
two times $t$ and $t'$, with $t'<t$, belong to the same domain
$D_u$ if we take the limits $t \to \infty$, $t' \to \infty$, with
the ratio $h_u(t')/h_u(t)$ finite and fixed to $e^{-\tau}$
($0 < \tau < \infty$). The $h_u$ are a family
of increasing functions indexed by a parameter $u$, such that,
if $w<u<v$ and the times $t,t'$ belong to
$D_u$, then $h_v(t')/h_v(t)=0$ and $h_w(t')/h_w(t)=1$
(a possible choice is $h_u(t)=exp(t^u)$, in which case $D_u$
is such that
$t'= t - \frac{t^{1-u}}{u} \tau$).
The domain $u=1$, with $h_1(t)=\exp(t)$,
 corresponds to the asymptotic regime where
FDT and TTI hold. We find convenient to express the fact that
$(t,t') \in D_u $ by the following diagram:
\begin{figure}
\centerline{\hbox{
\epsfig{figure=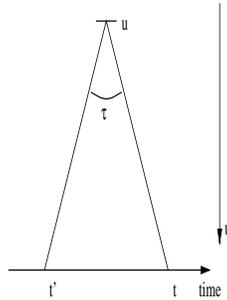,width=4cm,height=3cm,angle=-90}
}}
\caption{$(t,t') \in D_u $, with $h_u(t')/h_u(t)=e^{-\tau}$}
\end{figure}
It is then easy to show that, if we consider three times
$t' < s < t$, with $(s,t)  \in D_u$ and $(t',s)  \in D_v$,
then $(t',t)$ belong to $D_{min(u,v)}$ (see figure 5) which
is an ultrametric inequality. If for example $v > u$,
we have indeed $h_u(t')/h_u(s) = 1$, so $h_u(t')/h_u(t)
=h_u(s)/h_u(t)$.

\begin{figure}
\centerline{\hbox{
\epsfig{figure=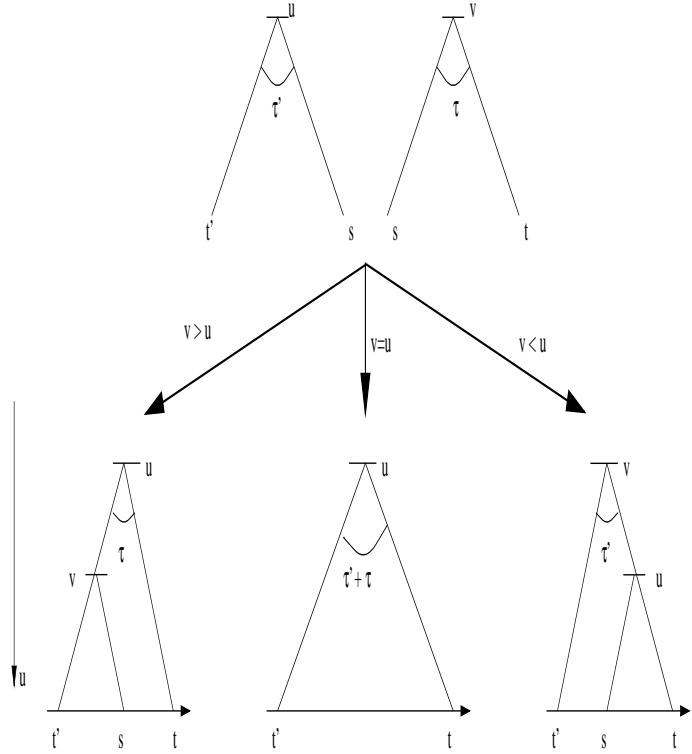,width=10cm,height=9cm,angle=-90}
}}
\caption{Ultrametric organizations of times}
\end{figure}

In each domain $D_u$, we assume the correlation and response to behave as:
\be
C(t,t')= C_u(\tau),~~~\  r(t,t')= \frac{d \ln (h_u(t'))}{dt'} r_u(\tau),
\ee
with a continuity condition: if $D_u$ and $D_v$ are neighbouring
domains, with $u < v$, then $C_u(0)=C_v(\infty)$.
Then it is possible to rewrite the equations (\ref{em})
(see appendix B for details), and to show \cite{kh2} that they
possess solutions obeying a generalized form of the fdt relation
called ``quasi-fdt'':
\be
x_u \frac{dC_u}{d\tau} = - Tr_u(\tau)
\ee

If we now consider three
times, $t' < s < t$ (see figure 5) the Ansatz implies a simple relation
between correlations at times $t,t'$, $t,s$ and $t',s$. When
$(t',s)$ and $(s,t)$ belong to two different domains we have:
\be
C(t,t')=min(C(t,s),C(s,t')),
\label{contact1}
\ee
and if they are in the same domain $D_u$, with
\be
\frac{h_u(t')}{h_u(s)} = e^{-\tau'} , \frac{h_u(s)}{h_u(t)} = e^{-\tau},
\mbox{then  } \frac{h_u(t')}{h_u(t)} = e^{-\tau'-\tau},
\ee
so that
\be
C(t,t')=C_u(\tau'+\tau)=j_u^{-1}(j_u(C(t,s))j_u(C(s,t'))),
\label{contact2}
\ee
where $C(t,s)=C_u(\tau)$, $C(s,t')=C_u(\tau')$, and $j_u(z)=
\exp(C_u^{-1}(z))$. Equations (\ref{contact1},\ref{contact2}) form the basis
of the formalism of triangular relations introduced in $\cite{cukusk}$,
and applied to the toy model in
\cite{cuglledou}. We provide in appendix C
 the solution for the
overlap function using this formalism.

Since the overlap function $Q_{t_w}(t,t')$ involves three times,
we are now looking for a function depending on the domains
$D_u$ and $D_{u'}$, $Q_{u,u'}(\tau,\tau')$,
where $(t_w,t) \in D_u$ and $(t_w,t') \in D_{u'}$
($h_u(t_w)/h_u(t)=e^{-\tau}$, $h_{u'}(t_w)/h_{u'}(t')=e^{-\tau'}$).

We rewrite in appendix B the equations (\ref{ove}) in this frame, and show
that they are solved by the following Ansatz:
\bea
&\hbox{If $u \neq u'$ (for example $u<u'$):}&\
Q_{t_w}(t,t') = C_u(\tau)=min (C(t,t_w),C(t',t_w)) \ . \ \ \\
&\hbox{ If $u=u'$:}&\  Q_{t_w}(t,t') = C_u(\tau' + \tau)=
j_u^{-1}(j_u(C(t,t_w))j_u(C(t',t_w))) \ .
\label{Qres}
\eea
This Ansatz can be easily understood in terms of the previously
introduced diagrams: at $t_w$ two ``time-sheets'' separate (see
figure 6) and two ultrametric systems appear, one for each replica.
We stress that this solution exists independently of the actual choice
of the disorder correlation, and therefore it is independent on the
precise solution of the aging dynamics: whatever the number of non
overlapping domains appearing in this solution, whatever the actual
solutions $C_u(\tau)$, there exists a  solution for the overlap
 function
in the aging regime which is
related to the correlation by (\ref{Qres}).

\begin{figure}
\centerline{\hbox{
\epsfig{figure=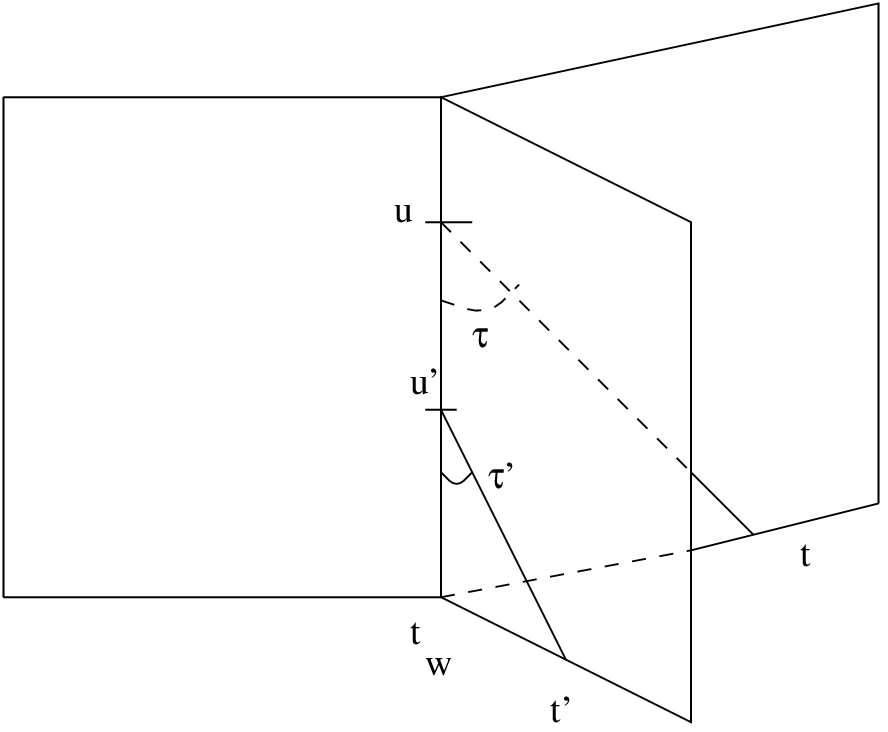,width=6cm,height=6cm,angle=-90}
\  \ \ \ \
\epsfig{figure=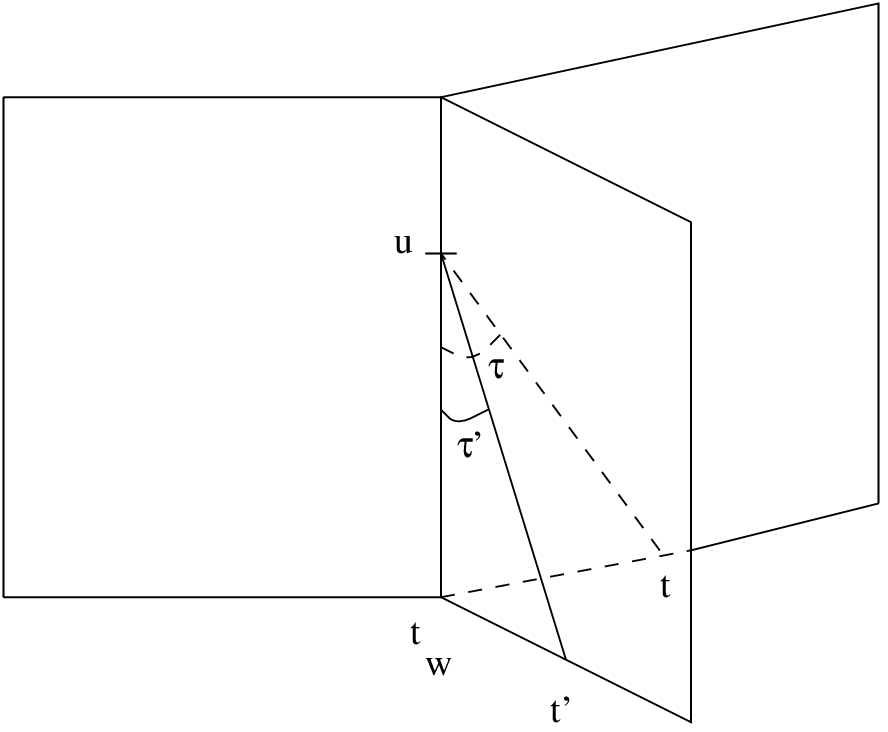,width=6cm,height=6cm,angle=-90}
}}
\caption{Two-sheets ultrametric structure for times bigger
than $t_w$}
\end{figure}

Depending on the model, the variable $u$ can be a priori
continuous, or discrete.
It was shown in \cite{franzmezard}, using the results from
\cite{kh2}, that for the
long-range model $u$ becomes a continuous variable.
For short-range models on the contrary, which exhibit
statically a one-step replica symmetry breaking, it has
been shown \cite{cukusk,cuglledou} that the aging dynamics is
solved by using a single time domain $D_{u^*}$
(beside the fdt  domain $D_1$).

To summarize, we have shown that, in the aging regime:
\begin{itemize}
\item for the long range model:
\be\label{llr}
Q_{t_w}(t,t') = \mbox{min} (C(t,t_w),C(t',t_w))
\ee
In particular, it is then clear that the long-time limit
of $Q_{t_w}(t,t')$ is $q_0$;
\item for the short-range model, there exists a function $j$ such that
\be\label{ssr}
Q_{t_w}(t,t') =j^{-1}(j(C(t,t_w))j(C(t',t_w))).
\ee
Since $j(q_0)=0$, and since $\lim_{t \to \infty} C(t,t_w) =
\lim_{t' \to \infty} C(t',t_w) =q_0$, we also have
$\lim_{t \to \infty} Q_{t_w}(t,t') =\lim_{t' \to \infty}
Q_{t_w}(t,t') = q_0$
\end{itemize}
In both cases, $\lim_{t_w \to \infty}\lim_{t \to \infty} Q_{t_w}(t,t)$ is
$q_0$, different from $\lim_{t \to \infty}
\lim_{t_w \to \infty}\ Q_{t_w}(t,t)$. Besides, no finite
waiting time is sufficient to give a higher limit than
$q_0$  for $\lim_{t \to \infty}Q_{t_w}(t,t)$: an increase
in the waiting time only slows down the dynamics, but has no
effect on the limiting values. No continuous
approach to equilibrium can thus be seen in this way.

{\it Note}: since the equations (\ref{em},\ref{ove}) are
causal, a numerical integration is available, as in
\cite{franzmezard}. Nevertheless, the integro-differential
character of these equations makes it difficult to reach
very long times (a huge amount of computer memory is needed).
Therefore, the numerical integration we were able to
realize, although fully compatible with the previous
study, were not conclusive enough to confirm it.

For the case of the p-spin spherical spin-glass model, an analytic
solution of the equations is also available: they are
solved by the Ansatz corresponding to the short-range model,
equation (\ref{ssr}). Indeed, we propose  for the aging regime
the Ansatz $Q_{t_w}(t,t')=Q(t_w/t,t_w/t')$, and we find
that the equation giving $Q_{t_w}(t,t')$ can be rewritten so
that the three times $t_w$, $t$ and $t'$ appear only through
the ratios $t_w/t$ and $t_w/t'$, and
that $Q_{t_w}(t,t')=\frac{1}{q} C(t,t_w)C(t',t_w) $
is solution of this equation.

We have thus shown that the overlap in the p-spin ($p \ge 3$)
spherical model exhibits aging in a
similar fashion as the correlation function, and decays
to zero for any finite $t_w$ ($q_0=0$ for this model). This
behaviour is thus very different from the $p=2$ or the
domain-growth case.

\subsection{Aging in traps}

A trap model was introduced in \cite{Bouchaud1} and developped
in \cite{BouDean} to reproduce off equilibrium dynamics
in glassy systems, and aging.
The model consists of $N$ traps with exponentially
distributed energy barriers. This distribution leads to
trapping time with infinite mean, and thus to aging.

In the simplest version, \cite{Bouchaud1} the basic
object is $\Pi _{1}(t,t_{w})$, probability that the system has not jumped
out of his trap
between $t_{w}$ and $t_{w}+t$. The overlap between two different
states is zero, and the self-overlap is $q_{EA}$. The correlation function is
then $q_{EA}\Pi _{1}(t,t_{w})$.

We now deal with two systems after $t_{w}$:
we introduce
$\Pi _{1} ^{(2)}(t,t',t_{w})$, probability that the first replica has not
jumped between $t_{w}$ and $t_{w}+t$, and that the second one
has not jumped between $t_{w}$ and $t_{w}+t'$.
The overlap $Q_{t_w}(t_{w}+t,t_{w}+t')$ is
then simply $q_{EA}\Pi _{1}^{(2)}(t,t',t_{w})$.

If the system is in trap
$\beta$ (of lifetime $\tau _{\beta}$) at time $t_{w}$, this probability is
$e^{-\frac{t}{\tau _{\beta}}} e^{-\frac{t'}{\tau _{\beta}}}$,
and thus we get ${\Pi}_{1} ^{(2)}(t,t',t_{w}) = {\Pi}_{1}(t+t',t_{w})$.
The overlap between the replicas is therefore:
\begin{equation}\label{q.c.rem}
Q_{t_{w}}(t_{w}+t,t_{w}+t')=C(t_{w}+t+t',t_{w})
\end{equation}

If we introduce a multi-layer tree, the only difference is that
we now have a set of $\Pi _{j}(t,t_{w})$ ($j=1,...,M$), probability
that the system has not jumped beyond the $j^{th}$ level of the tree
between $t_{w}$ and $t_{w}+t$. It is then clear that the equation
(\ref{q.c.rem}) is not changed, although the analytic expression for
the correlation function depends on the parameters of the tree.

For this particular model, the equilibrium relation
is in fact satisfied even for out of equilibrium dynamics,
because of the properties of the chosen exponential decay
from the traps. It is then clear that the overlap function
goes to zero for large times, for any finite $t_w$,
since $\lim_{t \rightarrow \infty} C(t_{w}+t,t_{w}) = 0$:
$S(t_w)=0$.
We have thus the same scenario as for the toy-model: for any
finite $t_w$, the overlap decays to its minimum allowed
value, while an infinite $t_w$ gives $q_{EA}$ as a limit.

\section{Conclusions}

In this paper, we have shown that the overlap between two
copies of a system, identical until a waiting time $t_w$,
and then totally independent, is a quantity of interest regarding
the geometry of phase space.
We have indeed studied this quantity for several models, and shown
that its decay is intimately related to the complexity of the
landscape and to the type of aging.
For simple systems, the long time limit of the overlap can
be put closer and closer to the equilibrium limit $q_{EA}$
by changing the time the replicas spend together. On the contrary,
for systems exhibiting a complex phase space, the limit of
the overlap is always the minimum value, i.e. the two replicas
are able to separate from each other, no matter how long
the waiting time is (if it stays finite). This difference
of behaviour can be quantitatively seen by a study of
the various long time limits of the overlap: for any system,
$\lim_{t\to \infty} \lim_{t_w \to \infty}Q_{t_w}(t_w+t,t_w+t)$
is $q_{EA}$, but the inverse order of limits,
$S_\infty=\lim_{t_w \to \infty} \lim_{t\to \infty}
Q_{t_w}(t_w+t,t_w+t)$ (and the behaviour of $S(t_w)=
\lim_{t\to \infty}Q_{t_w}(t_w+t,t_w+t)$), distinguishes between aging
in a ``simple'' phase space, and type II (spin glass) systems.

\noindent
{\bf Acknowledgements}

\noindent
We thank J.P. Bouchaud,
L. Cugliandolo, J. Kurchan, L. Laloux, P. Le Doussal, R. Monasson
and E. Vincent for
many interesting and helpful discussions on related topics.

\appendix

\section{Dynamical equations for the toy-model}

In the limit of infinite $N$, the dynamical equations
for the two-times correlation and response
functions (with $t > t'$) read \cite{franzmezard}:
\bea
{\partial r(t,t') \over \partial t}&=& -\mu r(t,t')
+\int_0^t ds  \ m(t,s)(r(t,t')-r(s,t'))\ ,
\\
{\partial C(t,t') \over \partial t} &=&-\mu C(t,t')
+ 2 \int_0^{t'} ds \  w(t,s) \ r(t',s)\nonumber\\
&+& \int_0^t ds \  m(t,s) \ (C(t,t')-C(s,t'))\ ,
\nonumber\\
{1 \over 2}{d C(t,t) \over  dt}&=& -\mu C(t,t)
+  2 \int_0^t ds  \ w(t,s) \ r(t,s) \nonumber\\
&+& \int_0^t ds \  m(t,s) \ (C(t,t)-C(s,t))\ +T \ ,
\label{em}
\eea
and the equations we obtain for $Q_{t_w}$ are:
\bea
\frac{\partial Q_{t_w}(t,t')}{\partial t} &=&
-\mu Q_{t_w}(t,t') + 2 \int_{0}^{t'} ds \  W_{t_w}(t,s) \ r(t',s) \\
&+& \int_{0}^{t} ds \  m(t,s) \ (Q_{t_w}(t,t') - Q_{t_w}(s,t')) \nonumber \\
\frac{1}{2} \frac{dQ_{t_w}(t,t)}{dt} &=& -\mu Q_{t_w}(t,t)
 + 2 \int_{0}^{t} ds \  W_{t_w}(t,s) \ r(t,s)  \nonumber \\
&+& \int_{0}^{t} ds \  m(t,s) \ (Q_{t_w}(t,t) - Q_{t_w}(s,t)) + T \theta
(t_{w}-t) \ ,
\label{ove}
\eea
with the same notation as \cite{franzmezard}:
\bea
w(t,t')&=&f'(b(t,t')), \ m(t,t')= 4 f''(b(t,t')) r(t,t') \\
 b(t,t')&=&C(t,t)+C(t',t')-2C(t,t') \
\eea
and:
\bea
W_{t_w}(t,t')&=&f'(B_{t_w}(t,t')) \\
B_{t_w}(t,t')&=&C(t,t)+C(t',t')-2Q_{t_w}(t,t') \ .
\eea

\section{Non overlapping time domains}

We show how to compute one of the integrals of equations
(\ref{em},\ref{ove}) in the frame of the Ansatz described
in \cite{franzmezard}, using the diagrams introduced in
figure 4. We apply this method to the overlap equation
and show that a similar Ansatz is solution.

For $(t',t) \in D_u$, we parametrize
\bea
b(t,t')= b_u(\tau),~~~\  r(t,t')= \frac{d \ln (h_u(t'))}{dt'} r_u(\tau),\\
m(t,t')=  \frac{d \ln (h_u(t'))}{dt'} m_u(\tau), \
w(t,t')= w_u(\tau).
\eea

The integral $\int_{t'}^t ds~ m(t,s)\ r(s,t')$, appearing
in (\ref{em}), has then three contributions (see figure 7):
\begin{itemize}
\item $(t',s) \in D_v$ ($v > u$), where $m(t,s)=m_u(\tau)
\frac{d\ln h_u(s)}{ds}$,
\item $t',t,s$ in the same domain $D_u$,
\item $(s,t) \in D_v$ ($v > u$), where $r(s,t')=r_u(\tau)$.
\end{itemize}

\begin{figure}
\centerline{\hbox{
\epsfig{figure=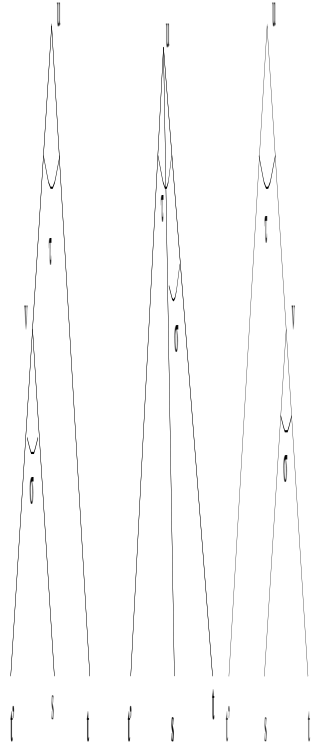,width=10cm,height=4cm,angle=-90}
}}
\caption{For $t' < s < t$, three regimes can be separated.}
\end{figure}

\bea
\int_{t'}^t ds~ m(t,s)\ r(s,t') = \frac{d\ln h_u(t')}{dt'}
& &\left( m_u(\tau) \sum_{v>u} \int_0^\infty\  d\sigma\  r_v(\sigma)
+ \int_0^\tau\  d\sigma\  m_u(\sigma)\  r_u(\tau - \sigma)
\right. \nonumber\\
& & \ \ +\left. r_u(\tau)  \sum_{v>u} \int_0^\infty\  d\sigma\  m_v(\sigma)
\right)
\eea

Separating in this way all different contributions in the integrals,
we obtain the following equation for $b_u(\tau)$:
\bea
0 &=&  b_u(\tau)\left[-\mu +  \sum_{v \leq u}
\int_0^{\infty} ds~ m_v(s) \right] + 2T -
\int_0^{\tau} ds~ m_u(s) ~ b_u(\tau-s)\nonumber\\
&-&  4~ w_u(\tau)~\sum_{v >u} \int_0^{\infty} ds~ r_v(s)
- \int_0^{\infty} ds~ \left[ m_u(\tau+s)~ b_u(s) + 4~w_u(\tau+s)~ r_u(s)
\right]\nonumber\\
&+& \sum_{v \geq u} \int_0^{\infty} ds~
\left[m_v(s)~  b_v(s) + 4~w_v(s)~ r_v(s)\right], \label{appbu}
\eea

The same method can be applied to the equation for
the overlap function $B_{t_w}(t,t')= B_{u,u'}(\tau,\tau')$. For
$ u,u' < 1$ (with $u \neq u'$) it reads:
\bea
0 &=&  B_{u,u'}(\tau,\tau')\left[-\mu +  \sum_{v \leq u}
\int_0^{\infty} ds~ m_v(s) \right] + 2T -
\int_0^{\tau} ds~ m_u(s) ~ B_{u,u'}(\tau-s,\tau')\nonumber\\
&-&  4~ w_u(\tau)~\sum_{u' > v >u} \int_0^{\infty} ds~ r_v(s)
 -4~ W_{u,u`}(\tau,\tau') ~ \sum_{v > u'} \int_0^{\infty} ds~ r_v(s)
\nonumber\\
&-& 4~ w_u(\tau) \int_{\tau'}^{\infty} ds~ r_{u'}(s)
- 4~ \int_0^{\tau'}ds~  W_{u,u`} (\tau, \tau'-s)~ r_{u'}(s)\nonumber\\
&-& \int_0^{\infty} ds~ \left[ m_u(\tau+s)~ b_u(s) + 4~w_u(\tau+s)~ r_u(s)
-  m_u(s)~ b_u(s) - 4~ w_u(s)~ r_u(s)\right]\nonumber\\
&+& \sum_{v > u} \int_0^{\infty} ds~
\left[m_v(s)~  b_v(s) + 4~w_v(s)~ r_v(s)\right]
\eea

If we insert in this equation the Ansatz
\be
B_{u,u'}(\tau,\tau') = b_u(\tau)\ \  (\mbox{if  } u < u' )
\ee
we reobtain equation (\ref{appbu}).
For $u=u'$, the equation is slightly different, and the Ansatz
\be
B_{u,u}(\tau,\tau') = b_u(\tau+\tau'),
\ee
together with the quasi-fdt,
again gives back equation (\ref{appbu}), evaluated at $\tau+\tau'$.

\section{Triangular relations}
We derive the solution for the overlap function using the formalism
of triangular relations \cite{cukusk,cuglledou}.
Neglecting time derivatives, and distinguishing aging and FDT regimes
in (\ref{em},\ref{ove}),
the final equation for the slow varying part (i.e. non fdt) of
$B_{t_w}(t,t')=C(t,t)+C(t',t')-2Q_{t_w}(t,t')$  reads
(with $q=2(\tilde{q}-q_1)$):
\bea
0&=& B_{t_w}(t,t')\left[~-\mu + \int_{0}^{t}ds~ m(t,s)\right] +
 2T -{2q\over T}~
[~W_{t_w}(t,t')-f'(q)~]\nonumber\\
&+& 4 \int_0^{t} ds~ w(t,s)~r(t,s)- 4 \int_0^{t_w} ds~w(t,s)~r(t',s)
-4 \int_{t_w}^{t'} ds~W_{t_w}(t,s) ~r(t',s)\nonumber\\
&+& \int_0^{t} ds~  m(t,s)~b(t,s)-\int_0^{t_w} ds~ m(t,s)~b(t',s)
- \int_{t_w}^{t} ds~ m(t,s)~ B_{t_w}(t',s)
\label{ebq}
\eea
where now all the times in the equations belong to the aging regime.

The approach of triangular relations measures the various
time domains directly in terms of the distance $b(t,t')$.
Indeed, there is a one to one correspondence between $b(t,t')$
(at large times) and the functions $b_u(\tau)$. More precisely,
if $t,t' \to \infty$ with $b(t,t')$ fixed to $B$, then $t,t'$ belong
to $D_u$, with $h_u(t')/h_u(t) = e^{-\tau}$, where $u$ and $\tau$
are fixed by $b_u(\tau)=B$. Then, the quasi-fdt can be expressed as
\be
r(t,t') = X[b(t,t')] {\partial b(t,t')\over \partial t'}
\ee
where $X[b(t,t')]=x_u/(2T)$.
Similarly, the ultrametric structure of time domains described
in (\ref{contact1},\ref{contact2}) can be written in a compact form as
a triangular relation\cite{cukusk,cuglledou}:
\be
b(t,t')=g(b(t,s),b(s,t')),
\ee
with $g(b,b')=\mbox{max}(b,b')$ when $b$ and $b'$ belong to different
domains (also named ``blobs''\cite{cukusk}),
and $g(b,b')=j^{-1}(j(b)j(b'))$ within the same domain.

The Ansatz concerning the function $B_{t_w}$ reads:
\be
B_{t_w}(t,t') = \gamma[ b(t,t_w),b(t',t_w)]
\ee
where $\gamma$ is a function to be determined.
Setting $b(t,t_w)=b_w$ , $b(t',t_w)=b_w'$, $b(t,t')=b$,
$b_0= 2(\tilde{q}- q_0)$ and
\be
F(b) = - \int_{b}^{q} ds~ X(s)
\ee
we obtain:
\bea
0 &=& \gamma(b_w,b_w')\left[-\mu + 4 \int_{b_0}^{q} ds~ f''(s)~X(s)\right]
 + 2T -{2q\over T}~
[f'[\gamma(b_w,b_w')]-f'(q)] \nonumber\\
&+& 4 f'(b_0)~F(b_0)+ 4 \int_{b_0}^{q}ds~ f''(s)~X(s)~s +
4 \int_{b_0}^{q} ds~f'(s)~X(s) \nonumber\\
&+& 4 \int_{b_0}^{b_w} ds~ f''(s)~ F[\bar{g}(b,s)]
-  4 \int_{b_0}^{b_w} ds~ f''(s)~X(s)~ \bar{g}(b,s)\nonumber\\
&+& 4\int_{b_w}^{b}ds~ F[\bar{g}(b,s)]~
f''[\gamma(b_w,\bar{g}(s,b_w))] \gamma'(b_w,\bar{g}(s,b_w))
\bar{g}'(s,b_w)\nonumber\\
&-&4  \int_{b_w}^{q}ds~ f''(s)~X(s)~\gamma[\bar{g}(s,b_w),b_w']
\label{eqpar}
\eea
where $\bar{g}$ is the reciprocal
function of $g$: given three times $t' < s < t$, in the limit
$t',s,t \to \infty$,
\be
b(t,t')=g(b(t,s),b(s,t')), \  b(s,t')=\bar{g}(b(t,s),b(t,t')).
\ee
Equation (\ref{eqpar}) is a functional
equation that gives $\gamma$ in terms of $f$, $X$ and $\bar{g}$.

Using the $g$-function described above, one can check from (\ref{eqpar})
that the solution is $\gamma(x,y)=g(x,y)$
(this means that the relation between the overlap
function $B_{t_w}(t,t')$ and the correlation functions
$b(t,t_w)$ and $b(t',t_w)$ is the same as the triangular
relation between the correlation functions).  This way, one gets back
the result for the overlap given in (\ref{llr},\ref{ssr}).

\end{document}